\documentclass{svproc}
\usepackage{url}
\usepackage{comment}

\usepackage{graphicx}
\usepackage{xcolor}
\begin{document}
\mainmatter              
\title{Thermodynamical analysis of QGP using effective PNJL model with Quasiparticle approach}
\titlerunning{Quasiparticle approach in Polyakov-Nambu-Jona-Lasinio Model}  
%
\author{Yogesh Kumar\inst{1} \and Himanshu Aggarwal\inst{1} \and R. S. Laishram
	\inst{1} \and
Poonam Jain\inst{2}$^{\textasteriskcentered}$ \and Om Prakash\inst{2} \and Sanjay Tyagi\inst{2} \and Pargin Bangotra\inst{3} \and Vinod Kumar\inst{4} \and Yagyadatta Goswami\inst{5} \and M. K. Sahu\inst{5}}
\authorrunning{Y. Kumar et al.} 
%
\tocauthor{Poonam Jain, Himanshu Aggarwal, R. S. Laishram, Yogesh Kumar, Pargin Bangotra, Vinod Kumar, Yagyadatta Goswami and M. K. Sahu}
\institute{$^1$Department of Physics, Hansraj College, University of Delhi, Malka Ganj, New Delhi-110007, India
	\\$^2$Department of Physics, Sri Aurobindo College, University of Delhi, Malviya Nagar, New Delhi-110017, India
	\\
	$^3$Department of Physics, Netaji Subhas University of Technology, Dwarka, Delhi, India
	\\
	$^4$Department of Physics, University of Lucknow, Lucknow-226007, U.P., India \\
	$^5$ Department of Electronics, Hansraj College, University of Delhi, Malka Ganj, New Delhi-110007, India \\
\email{$^{\textasteriskcentered}$Corresponding author: pjain$\_$phy@aurobindo.du.ac.in}
}

\maketitle              

\begin{abstract}
We study the thermodynamics of the quark-gluon plasma using an effective 2-flavor Polyakov–Nambu–Jona‑Lasinio (PNJL) model extended by a quasiparticle description for quarks and gluons, incorporating temperature dependent quark masses within the PNJL framework. Two variants, Quasiparticle Model‑I and Quasiparticle Model‑II, are implemented to investigate bulk thermodynamic observables such as pressure, energy density, entropy density, specific heat, and the speed of sound. The combined framework yields a robust baseline for the description of hot QGP dynamics in the high temperature regime at vanishing chemical potential and zero magnetic field. Systematic comparison with lattice QCD results shows an excellent agreement and clear improvement over conventional PNJL implementations. We observe that both variants complement each other, offering mutually consistent insight into quasiparticle mass effects and medium response in the deconfined phase. This mutual consistency validates the physical foundation of the overall quasiparticle mechanism, reinforcing the credibility of the calculated Equation of State. Finally, the quasiparticle model extension improves PNJL from a descriptive tool to a more qualitative phenomenological approach, enabling an improved description of the strong interacting quark-gluon plasma.

\end{abstract}
\section{Introduction}
In the first few microseconds after the Big Bang, the universe was a super heated soup of elementary particles known as the quark-gluon plasma (QGP) \cite{c1,c2}. This primordial state represents a key phase in the life of Quantum Chromodynamics (QCD), the theory of the strong force. Today, QCD's influence is seen in two non-perturbative phenomena: color confinement, which perpetually traps quarks and gluons within composite particles called hadrons \cite{c1,c3}, and spontaneous chiral symmetry breaking, the mechanism responsible for generating the dominant portion of hadron mass \cite{c4,c5,c6}. The transition from the primordial QGP to this familiar hadronic matter, therefore, represents one of the most profound phase changes in cosmic history. Recreating and studying this phase transition is a central goal of high energy physics, providing crucial insights into the fundamental properties of matter under extreme conditions.

Experimental facilities like the Large Hadron Collider (LHC) at CERN and the Relativistic Heavy Ion Collider (RHIC) at BNL, collide heavy nuclei at near light speeds to shortly recreate these extreme temperatures and densities \cite{c2,c3,c7,c8}. The aim is to probe the deconfining phase transition, where quarks and gluons are liberated and the simultaneous chiral symmetry restoration (CSR), where they shed their dynamically generated mass \cite{c1,c2,c3,c4,c5}. This transition is not a sharp event but a continuous crossover, with the Polyakov loop acting as an indicator for deconfinement and the chiral condensate for CSR \cite{c2,c3,c9}.

Despite the considerable amount of experimental data, making accurate theoretical predictions about the QGP remains a formidable challenge. The strong coupling of QCD near the transition makes conventional methods like perturbative QCD (pQCD) and Hard Thermal Loop (HTL) resummation unreliable \cite{c3,c10,c11,c12}. Moreover, while lattice QCD (LQCD) is a powerful tool at zero chemical potential, the "fermion sign problem" makes it computationally intractable at non-zero values, leaving a significant gap in our understanding of the QCD phase diagram \cite{c2,c4,c13,c14}. This "non-perturbative gap" has underscored the necessity of robust phenomenological models. Early models were often thermodynamically inconsistent or provided a wide range of predictions for the Critical End Point (CEP), further complicating the picture \cite{c1,c2,c7}.

To bridge this gap, the Nambu–Jona-Lasinio (NJL) model was proposed by Nambu and Jona-Lasinio in 1961 \cite{c6,c14} to describe chiral symmetry breaking \cite{c6,c15,c16}. Its extension, the Polyakov–Nambu–Jona-Lasinio (PNJL) model \cite{c17}, incorporates deconfinement dynamics through the Polyakov loop \cite{c4,c9,c17,c18}. This development created a self consistent framework that successfully predicted the simultaneous occurrence of chiral restoration and deconfinement, in good agreement with lattice simulations. The predictive capability of the PNJL Model is evident in its close agreement with numerous lattice QCD results, including pressure and energy density \cite{c2,c3,c4,c9,c18}.

However, as an effective description, PNJL exhibits limitations in quantitative accuracy near the transition region. This has led to several refinements including improved Polyakov loop potentials and analyses of higher order fluctuations related to heavy ion observables and the critical end point (CEP) \cite{R1,R2}. Several extensions implementing finite density effects, finite volume corrections and vector interactions have further enriched its phase structure \cite{R3}. Comparisons with functional approaches such as the Functional Renormalization Group (FRG) provide important benchmarks for assuring its validation \cite{R4}. These developments indicate that PNJL captures essential features of QCD thermodynamics but achieving quantitative agreement with first principle results remains an open challenge.
\\
To address these limitations, alternative approaches have been developed to more directly reproduce the thermodynamic properties of QCD matter. Quasiparticle models (QPMs) provide one such framework, pioneered by seminal works from QPM-I \cite{c7} and QPM-II \cite{c10,c11}. These models simplify the complex system of interacting quarks and gluons by treating them as a gas of effective, massive particles. The crucial difference lies in how these quasiparticles are defined. Rather than being dynamically generated from a Lagrangian, the effective temperature dependent masses are introduced phenomenologically, often with a functional form that is constrained by fitting to lattice data or HTL results \cite{c7,c19}. This approach has been particularly successful in describing the Equation of State (EoS), including the softening near the transition region reflected in the speed of sound \cite{c2,c7,c19}.
\\
Modern quasiparticle formulations focus on further improving thermodynamic consistency and reducing phenomenological inputs. Early studies \cite{R5} emphasize thermodynamic self consistency conditions, while later works \cite{R6,R7} incorporate medium dependent dispersion relations within transport frameworks. Extensions to finite baryon density, along with improved agreement with lattice QCD \cite{R8}, further enhance their reliability. This leads to a more constrained and QCD consistent description of the quark-gluon plasma.
\\
Following these developments, the present work combines PNJL-based confinement dynamics with temperature dependent quark masses within a quasiparticle framework, leading to an effective PNJL description. The framework is implemented at vanishing chemical potential ($\mu_0=0$) and zero magnetic field ($B=0$) to investigate the thermodynamic properties of the Equation of State with an extended approach. This formulation aims to provide improved insight into the interaction between confinement effects and medium modified quasiparticle excitations in the quark-gluon plasma. Accordingly, the analysis is performed within a 2-flavor setup, which serves as a baseline for isolating light-quark dynamics. Given the limited availability of 2-flavor lattice QCD data, comparisons are made with 2+1-flavor results to assess qualitative trends and to estimate the impact of the missing strange quark degree of freedom. 
\\
The manuscript is organized as follows: In section \ref{Section:2}, we briefly introduce the quasiparticle approach within the PNJL model and further evaluate the concerning thermodynamic parameters to define the Equation of State (EoS) in section \ref{Section:3}. We analyse the results in section \ref{Section:4}. Finally, discuss the conclusion in section \ref{Section:5}.

\section{A brief overview of QPM in PNJL Model}
\label{Section:2}
The PNJL model \cite{c17}, a well-established effective theory of QCD, serves as the foundation for our investigation. This framework has proven highly effective in describing the crossover transition to the QGP and reproducing a wide range of thermodynamic observables. We build upon the foundation of 2-flavor PNJL model by adopting a quasiparticle based approach. 
\\
The temperature dependent masses of these quasiparticle employed in this work follow the formulations of QPM-I \cite{c7}. This approach considers the thermal mass $m_q$ that arises due to the interactions among quarks and gluons within the surrounding matter of the medium. Whereas, its refined extension QPM-II \cite{c10,c11}, incorporates the current quark mass with the thermal mass in the system.
\\
In this manuscript, these models are used to define the dynamics of the QGP using quasiparticle approach. Within this formalism, the effective mass of quasiparticles, named as QPM-II, which incorporates both current and thermal contributions is expressed as \cite{c8,c11,c20,c21}:
\begin{eqnarray}
	\label{Eq:2}
	m_{eff}^2&=&(m_c+m_q)^2+m_q^2,
\end{eqnarray}
where $m_c$ represents the current quark mass and $m_q$ denotes the thermal quark mass as QPM-I. The thermal mass is modeled as \cite{c5,c22,c23}:
\begin{eqnarray}
	\label{Eq:3}
	m_q^2(T)&=&(4\pi\alpha_s)\gamma_qT^2,
\end{eqnarray}
with $\alpha_s$ is the strong coupling constant and $\gamma_q$ is a flow parameter used to predict the stability of the QGP flow and order of phase transition. The coupling constant itself depends on the QCD scale parameter $\Lambda_{QCD}$ and is parameterized as \cite{c8}:
\begin{eqnarray}
	\alpha_s&=&\frac{4}{(11N_c-2N_f)\ln\left(1+\frac{k^2}{\Lambda_{QCD}^2}\right)},\nonumber
\end{eqnarray}
where $N_c=3$ is the number of colors, $N_f=2$ is the number of flavors, and $k$ is a scale factor related to the thermal environment \cite{c22,c23}:
\begin{eqnarray}
	k&=&\left[\frac{\gamma N^{\frac{1}{3}}T^2\Lambda_{QCD}^2}{2}\right]^{\frac{1}{4}}.\nonumber
\end{eqnarray}
Where, the parameterization factors, $\gamma=\sqrt{2[\gamma_q^{-2}+\gamma_g^{-2}]}$ and $N=16\pi/(11N_c-2N_f)$.
The parameters $m_c$, $\Lambda_{QCD}$, and numerical coefficients $\gamma_g$ and $\gamma_q$ are provided in Table \ref{Tab:1}.
\\
\begin{table}
	\centering
	\caption{Current quark mass and other parameters in effective mass $m_{eff}$.}
	\setlength{\tabcolsep}{15pt}
	\renewcommand{\arraystretch}{1.7}
	\begin{tabular}{|c|c|c|c|c|}
		\hline
		$m_c$[MeV]  &   $\Lambda_{QCD}$[MeV]  &  $\gamma_q$  &  $\gamma_g$  \\
		\hline
		10 & 150 & 1/6 & 6 or 8$\gamma_q$ \\
		\hline
	\end{tabular}
	\label{Tab:1}
\end{table}

The overall thermodynamic potential for 2-flavor system is then constructed with the quasiparticle contributions and the effective potential of the Polyakov loop. Under the conditions of zero magnetic field, this potential is formally expressed as follows \cite{c1,c5}:

\begin{eqnarray}
	\label{Eq:1}
	\Omega&=&\mathcal{U}(\phi,\overline{\phi},T)+\frac{(m-m_c)^2}{2G}-2N_f\int\frac{d^3p}{8\pi^3}\left\{3E_p\theta(\Lambda^2-p^2)+\right.\nonumber\\
	&& \left.T\ln [1+3\phi e^{-(E_p-\mu_0)/T}+3\overline{\phi}e^{-2(E_p-\mu_0)/T}+e^{-3(E_p-\mu_0)/T}]\right.\nonumber\\
	&& \left.+T\ln [1+3\overline{\phi}e^{-(E_p+\mu_0)/T}+3\phi e^{-2(E_p+\mu_0)/T}+e^{-3(E_p+\mu_0)/T}]\right\},
\end{eqnarray}

where, for a given system the quasiparticle energy $E_p=\sqrt{p^2+m^2}$, $\mu_0$ is the quark chemical potential fixed at zero, and the fields $\phi$ and $\overline{\phi}$ represent the Polyakov loop and its conjugate, respectively. For the cases of quasiparticle model as QPM-I and QPM-II, the mass ($m$) would be translated into thermal mass $m_q$ and effective mass $m_{eff}$, respectively.

Whereas, the parameters $\Lambda$ and $G$ define the regularization scale and the strength of the scalar interaction term appearing in the thermodynamic potential. These parameters are taken from standard PNJL parameter sets and fixed at values 651 MeV and $10.08\times10^{-6}$ MeV$^{-2}$, respectively. In the present framework, they serve as effective inputs controlling the vacuum contribution and residual interaction structure. Variations in these parameters primarily influence the quantitative scale of the thermodynamic observables without modifying the overall qualitative behaviour of the EoS.

The first term in Eq. \ref{Eq:1}, $\mathcal{U}(\phi, \overline{\phi},T)$ corresponds to the Polyakov loop potential, which governs the confinement–deconfinement dynamics. In the commonly used polynomial form \cite{c17}, it is written as:
\begin{eqnarray}
	\label{Eq:4}
	\frac{\mathcal{U}(\phi,\overline{\phi},T)}{T^4}&=&-\frac{b_2(T)}{2}\overline{\phi}\phi-\frac{b_3}{6}(\phi^3+\overline{\phi}^3)+\frac{b_4}{4}(\phi\overline{\phi})^2.
\end{eqnarray}
In the above expression, $b_2(T)$ is expressed as:
\begin{eqnarray}
	\label{Eq:5}
	b_2(T)&=&a_0+a_1\left(\frac{T_0}{T}\right)+a_2\left(\frac{T_0}{T}\right)^2+a_3\left(\frac{T_0}{T}\right)^3.
\end{eqnarray}
Here, $T_0$ is the pure-gauge critical temperature and  $a_i$, $b_i$, are the parameters fitted to lattice QCD data \cite{c24,R9}, provided in Table \ref{Tab:2}.
The value of $T_0$ is taken to be 190 MeV for the transition temperature $T_c$ to be near 170 MeV. Additionally, the Polyakov loop potential in Eq. \ref{Eq:4} is taken from pure SU(3) LQCD, as standard in PNJL models. While the presence of quarks through QPM models do not modify the potential itself, it affect the Polyakov loop expectation value at the mean-field level through their coupling in the distribution functions. This mechanism suppresses color degrees of freedom at lower temperatures and restores gluonic contributions at higher temperatures. 
\\
\begin{table}
	\label{Tab:2}
	\centering
	\caption{Parameters used in the Polyakov loop potential in Eqs. \ref{Eq:4} and \ref{Eq:5} are taken from Ref. \cite{c1}.}
	\setlength{\tabcolsep}{15pt}
	\renewcommand{\arraystretch}{1.7}
	\begin{tabular}{|c|c|c|c|c|c|}
		\hline
		$a_0$  &  $a_1$  &  $a_2$  & $a_3$  &  $b_3$ & $b_4$  \\
		\hline
		6.75 & -1.95 & 2.625 & -7.44 & 0.75 & 7.5 \\

		\hline
	\end{tabular}
\end{table}

\section{Thermodynamic Observables  with Quasiparticle Approach}
\label{Section:3}
To formalize and obtain the thermodynamic parameters using the thermodynamic potential($\Omega$), two parameters would be required to determine, i.e., $\phi$ and $\overline{\phi}$. The appropriate method for finding these parameters is to minimize $\Omega$ with respect to the fields $\phi$ and $\overline{\phi}$.
\begin{eqnarray}
	\frac{\partial\Omega}{\partial\phi}=0, \frac{\partial\Omega}{\partial\overline{\phi}}=0~.
\end{eqnarray}
Solving these coupled equations would allow us to determine the observables. At zero chemical potential $\mu_{0}$, both the Polyakov loop and its conjugate would yield the same value.
Then, the transition temperature $T_c$ can be calculated by analyzing the behaviour of $d\phi/dT$ and identifying the temperature at which it reaches its peak value. By determining these quantities, the thermodynamic parameters can be formalized.
First, we determine pressure, which can be defined as a function of temperature for zero chemical potential $\mu_0$:
\begin{eqnarray}
	P(T,\mu_0=0)&=&-\Omega(T,\mu_0=0)~,
\end{eqnarray}
where the vacuum components are removed from the thermodynamic potential to obtain vanishing pressure and other thermodynamic quantities at $T=0$ and $\mu_0=0$. Next, the entropy density can be calculated as:
\begin{eqnarray}
	s(T,\mu_0=0)&=&-\frac{\partial\Omega(T,\mu_0=0)}{\partial T}~.
\end{eqnarray}
Given pressure and entropy, energy density can be defined as:
\begin{eqnarray}
	\epsilon(T,\mu_0=0)&=&-P+Ts=-T\frac{\partial\Omega(T,\mu_0=0)}{\partial T}+\Omega(T,\mu_0=0)~.
\end{eqnarray}

The first derivative of energy density with respect to temperature is termed as specific heat $C_V$. It is defined as:
\begin{eqnarray}
	C_V&=&\frac{\partial\epsilon}{\partial T}
	=-T\frac{\partial^2\Omega(T,\mu_0=0)}{\partial T^2}~.
\end{eqnarray}
Subsequently, the squared speed of sound can be defined by pressure and energy density at constant entropy as:
\begin{eqnarray}
	v^2_s&=&\left.\frac{\partial P}{\partial\epsilon}\right|_{s}
	=\left.\frac{\partial P}{\partial T}\right/\frac{\partial \epsilon}{\partial T}
	=\left.\frac{\partial\Omega(T,\mu_0=0)}{\partial T}\right/T\frac{\partial^2\Omega(T,\mu_0=0)}{\partial T^2}.
\end{eqnarray}
With the help of these thermodynamic parameters at fix value of zero chemical potential and zero magnetic field, the results are explained in the next section 4 using QPM-I and QPM-II.

\section{Results}
\label{Section:4}
We present the behaviour of various thermodynamic quantities as functions of scaled temperature, $T/T_c$. The critical temperature is observed as $T_c=168.8$ MeV with the pure gauge critical temperature fixed at $T_0=190$ MeV. This value of $T_c$ is consistent with 2-flavor lattice QCD data, which yields $T_c=173 \pm 8$ MeV. Unlike other PNJL-based studies, the present work implements temperature dependent quasiparticle masses within the effective PNJL framework instead of masses generated through the chiral condensate. This choice was motivated by the effectiveness of quasiparticle descriptions in reproducing QCD thermodynamics near and above the transition temperature, where medium induced effects dominate.
Within this setup, we analyze the differences arising from the use of two distinct types of quasiparticle models: QPM-I and QPM-II. These two models are then used in describing the behaviour of various thermodynamic quantities.
Furthermore, our results are compared with the recent lattice QCD data \cite{c25,c27,c28} and other theoretical models \cite{c29,c30,c31}, both of which correspond to the 2+1-flavor regime. This choice is necessitated by the limited availability of 2-flavor data. Accordingly, these comparisons are intended to assess qualitative consistency and general trends rather than exact quantitative agreement, since 2+1-flavor results serve as the current standard reference in the field.

\begin{figure*}[!th]
	\centering
	\resizebox{9cm}{!}{\includegraphics{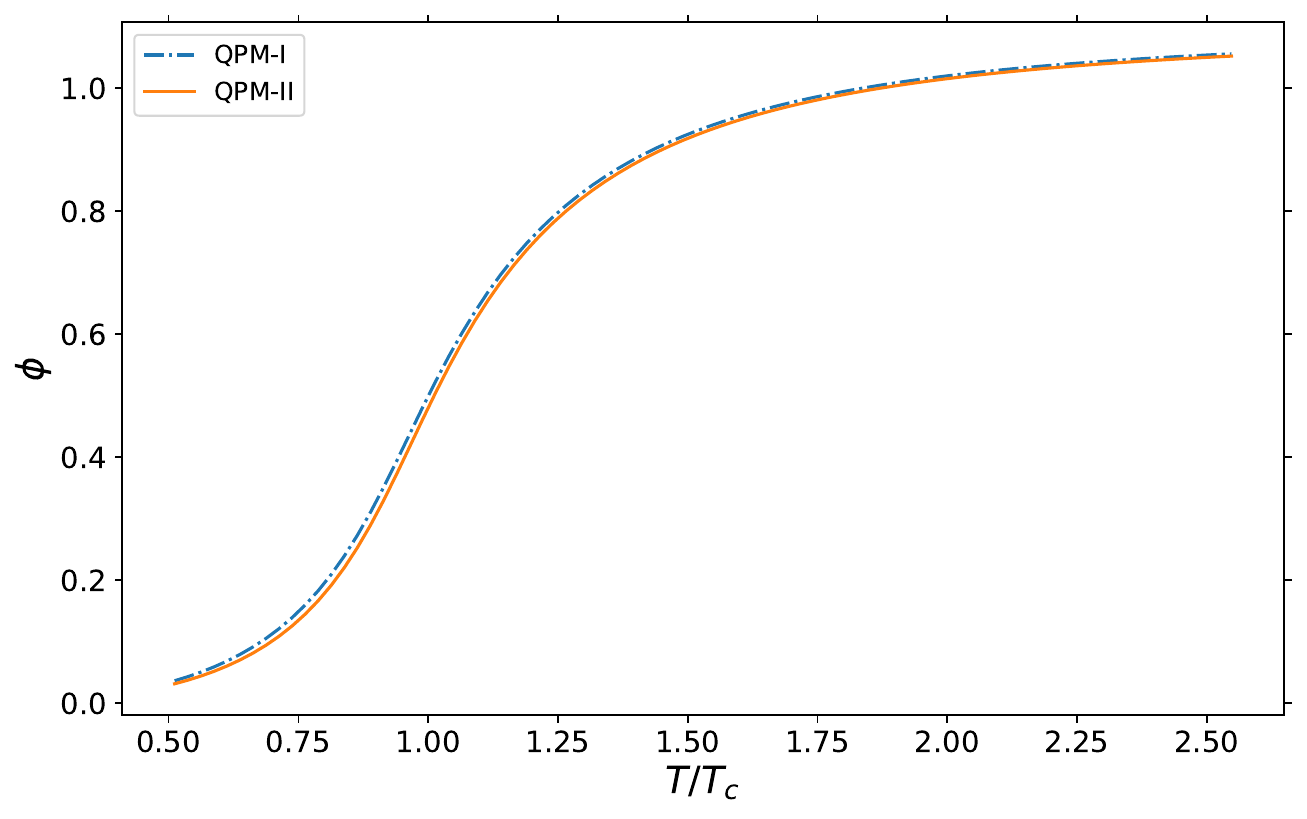}}
	\caption{Polyakov loop as a function of scaled temperature at $\mu_0$=0 for QPM-I and QPM-II are shown, in 2-flavor regime}.
	\label{Phi}
\end{figure*}

The Polyakov loop $\phi$, serves as an approximate order parameter (or indicator) for the confinement–deconfinement transition. At $\mu_0$=0, it follows that $\phi$=$\overline{\phi}$. Fig. \ref{Phi} shows the variation of $\phi$ with $T/T_c$. Both QPM-I and QPM-II reproduce the expected confinement–deconfinement transition: the Polyakov loop rises from nearly zero at low temperature to unity in the deconfined regime, with the QPM-I curve lying slightly above that of QPM-II. Within the PNJL framework, this behaviour reflects the role of $\phi$ in modulating the gluonic background. In contrast to pure quasiparticle models, this leads to a suppression of thermodynamic excitations in the confined phase ($\phi \approx0$). As the temperature increases, the gradual approach of $\phi$ toward unity signals the onset of deconfinement and the corresponding restoration of gluonic contributions.
The two formulations yield nearly identical results with only minor deviations around the crossover region and thus indicates that the quasiparticle extension retains the essential deconfinement dynamics inherent to the PNJL framework.
\begin{figure*}[!th]
	\resizebox{12.5cm}{!}{\includegraphics{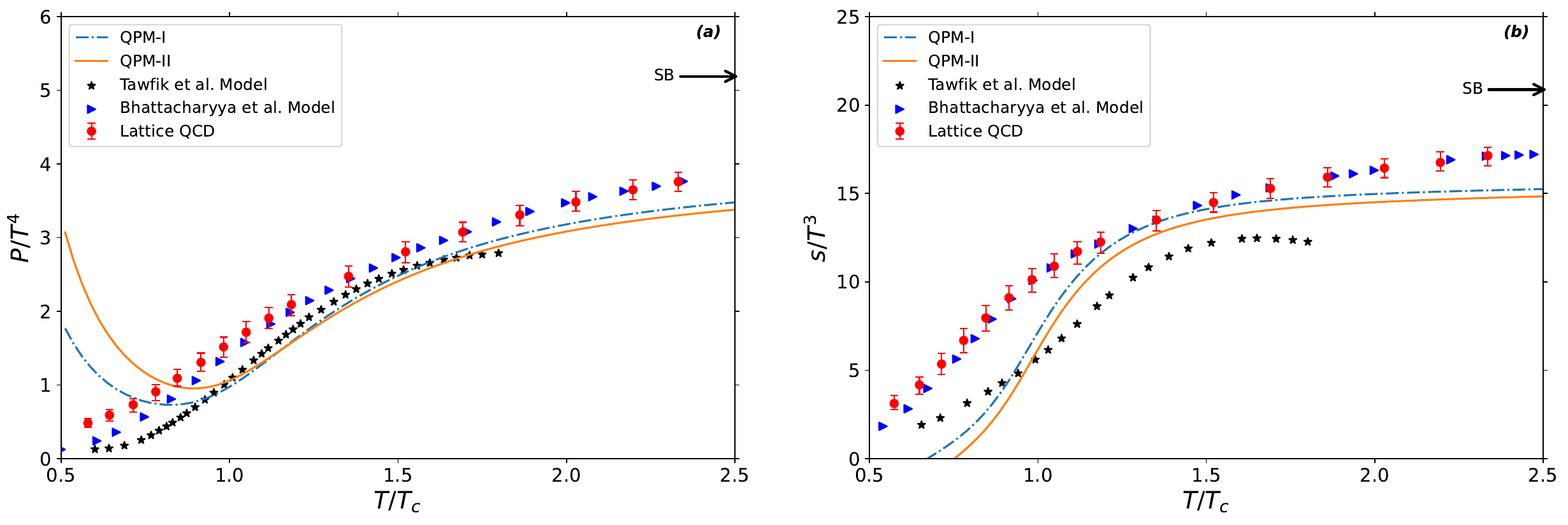}}
	\caption{Variation of (a) scaled pressure $P$, (b) scaled entropy density $s$ as a function of scaled temperature with $\mu_0$=0 for QPM-I and QPM-II are shown, in 2-flavor regime. The results are compared with the 2+1-flavor lattice QCD data \cite{c25}, and the results of Bhattacharyya et al. \cite{c30} and Tawfik et al. \cite{c31}, which are also within the 2+1-flavor setup. The Stefan–Boltzmann (SB) limit is defined as $\frac{P_{SB}}{T^4}=(N_c^2-1)\frac{\pi^2}{45}+N_cN_f\frac{7\pi^2}{180}$ and $\frac{s_{SB}}{T^3}=4\frac{P_{SB}}{T^4}$, evaluated for $N_f=2+1$ and consistent with the lattice data.}
	\label{Therm1}
\end{figure*}
\begin{figure*}[!th]
	\centering
	\resizebox{12.5cm}{!}{\includegraphics{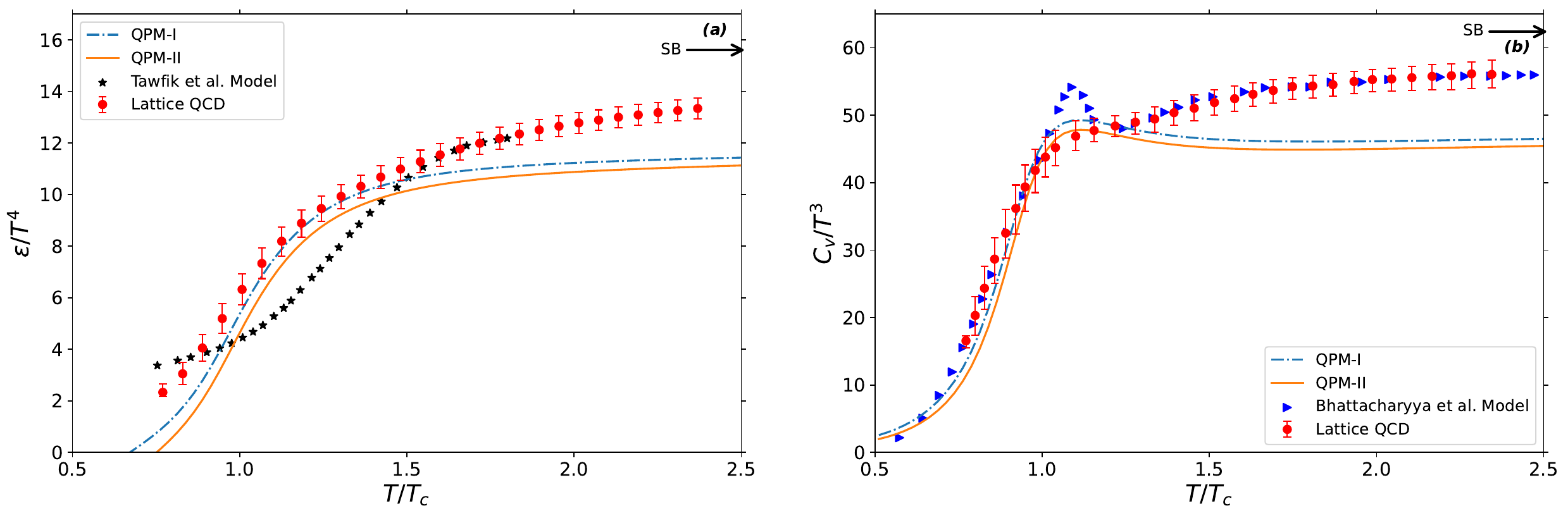}}
	\caption{Variation of (a) scaled energy density $\epsilon$ and (b) scaled specific heat $C_V$ as a function of scaled temperature with $\mu_0$=0 for QPM-I and QPM-II are shown, in 2-flavor regime. The results are compared with the 2+1-flavor lattice QCD data \cite{c25}, and the results of Bhattacharyya et al. \cite{c30} and Tawfik et al. \cite{c31}, which are also within the 2+1-flavor setup. The Stefan–Boltzmann (SB) limit is defined as $\frac{\epsilon_{SB}}{T^4}=3\frac{P_{SB}}{T^4}$ and $\frac{C_v^{SB}}{T^3}=12\frac{P_{SB}}{T^4}$, evaluted for $N_f=2+1$ and consistent with the lattice data.}
	\label{Therm2}
\end{figure*}

Fig. \ref{Therm1} (a) compares our results of scaled pressure with lattice QCD data \cite{c25} and the results of Bhattacharyya et al. \cite{c30} and Tawfik et al. \cite{c31}. For $T/T_c<1.15$, the QPM-II curve lies above the QPM-I curve and decreases to a dip around $T/T_c\approx0.8$, after which it rises sharply in the range $0.9\leq T/T_c\leq1.0$. Similar behaviour is shown by the QPM-II curve with a dip around $T/T_c\approx0.75$, followed by a sudden increase near $T/T_c\approx0.9$. These features reflect the confinement–deconfinement transition and highlight the strong role of thermal interactions among quasiparticles near the critical region. In the interval $1.15\leq T/T_c\leq1.2$, the two curves converge, while for $T/T_c>1.2$, the QPM-I curve remains slightly above that of QPM-II. At higher $T$, both masses tend toward perturbative behaviour and the curves converge, as expected from quasiparticle EoS systematics
and lattice‑constrained PNJL fits. Overall, both QPM‑I and QPM‑II agree closely with lattice QCD \cite{c25} simultaneously, and the works of Bhattacharyya et al. \cite{c30} and Tawfik et al. \cite{c31}, particularly for
$0.95\leq T/T_c\leq1.8$, a region characterized by a sharp increase.

In Fig. \ref{Therm1} (b), the scaled entropy density for QPM‑I remains above QPM‑II throughout $0.5\leq T/T_c\leq2.5$, with both curves expected to converge at higher temperatures and saturate close to the Stefan-Boltzmann limit. Both curves depicts qualitative agreement with the lattice QCD \cite{c25} and the Bhattacharyya et al. work \cite{c30}, particularly in the region $1.2<T/T_c<1.7$, indicating that the quasiparticle–PNJL framework reliably captures the thermodynamics of the transition. Their overall trend also bears some resemblance to the pattern observed in the Tawfik et al. work \cite{c31}.

The Fig. \ref{Therm2} (a) shows the scaled energy, which exhibits a rapid rise around $T_c$, consistent with the sudden increase in active degrees of freedom during the crossover. The QPM-I curve remains slightly above QPM-II across the studied temperature range, although both align very well with lattice QCD data \cite{c25}, again reflecting the combined effect of deconfinement and strong medium dependence of quasiparticle interactions.

\begin{figure*}[!th]
	\centering
	\resizebox{12.5cm}{!}{\includegraphics{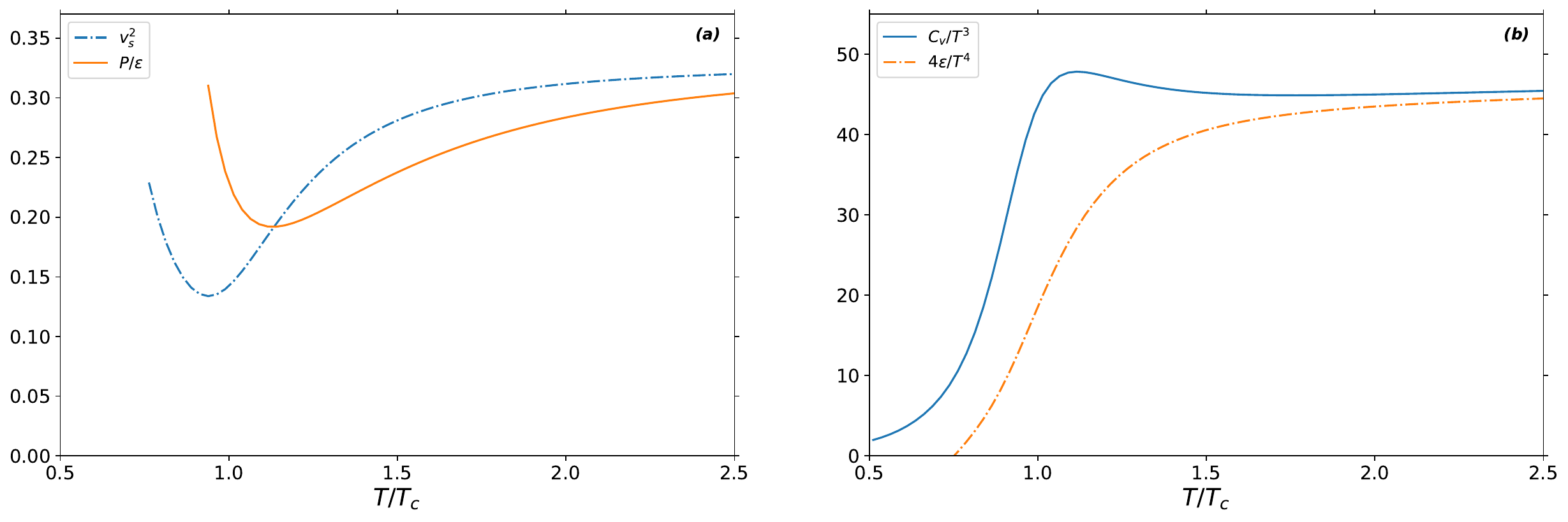}}
	\caption{Comparisons between: (a) speed of sound squared and $P/\epsilon$ (a), and (b) scaled energy density $\epsilon$ and scaled specific heat $C_V$ as a function of scaled temperature with $\mu_0$=0 for QPM-II are shown, in 2-flavor regime.}
	\label{Therm4}
\end{figure*}

In Fig. \ref{Therm2} (b), we can observe the behaviour of scaled specific heat with temperature. In this figure, a slight peak is observed near $T/T_c\approx1.0$ for both the QPM-I and QPM-II curves. This peak is characteristic of the crossover transition, where fluctuations are enhanced and the system undergoes rapid changes in its thermodynamic state. This prominent feature also underscores the thermodynamic sensitivity of the medium in the vicinity of critical temperature. Although these peaks are less pronounced than in the model of Bhattacharyya et al. \cite{c30}, the results exhibit strong resemblance to both, that model and lattice data up to $T/T_c\approx1.0$. The specific heat also approaches the conformality limit at higher temperatures (see also Fig. \ref{Therm4}(b)). Across the full temperature range, the QPM-I curve lies slightly above the QPM-II curve.

\begin{figure*}[!th]
	\centering
	\resizebox{12.5cm}{!}{\includegraphics{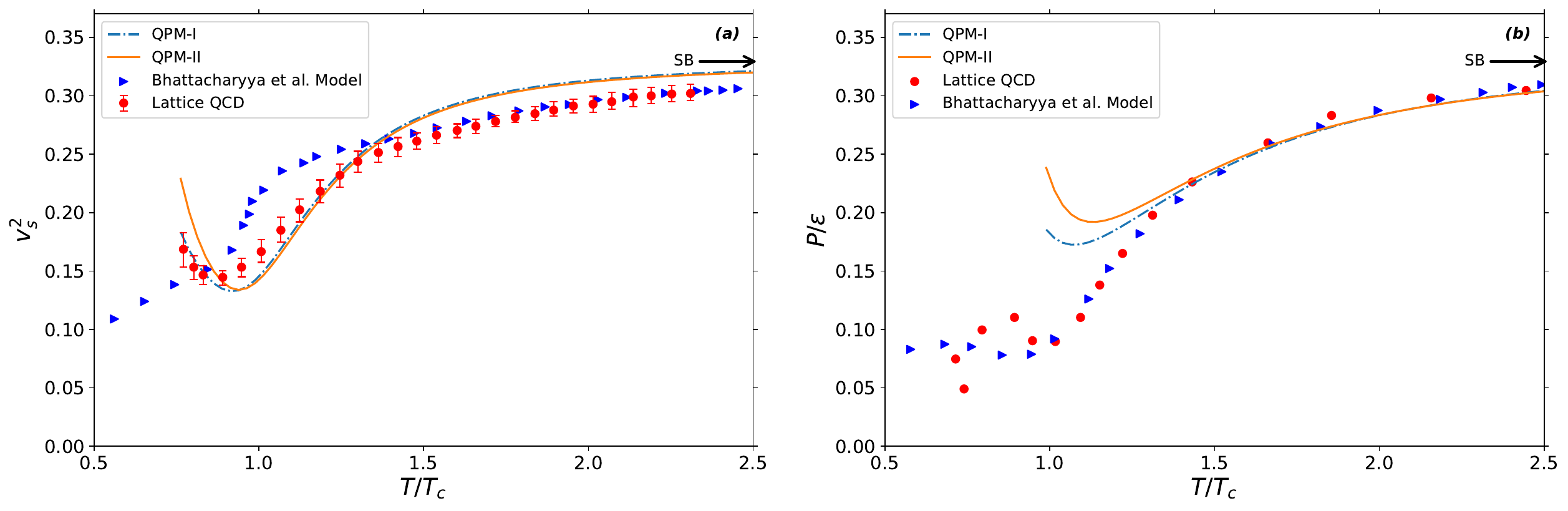}}
	\caption{Variation of (a) speed of sound squared $v_s^2$ and (b) $P/\epsilon$ as a function of scaled temperature with $\mu_0$=0 for QPM-I and QPM-II are shown, in 2-flavor regime. The results are compared with the 2+1-flavor lattice QCD data \cite{c25,c27,c28} (in which the LQCD data for $\frac{P}{\epsilon}$ corresponds to its mean value), and the results of Bhattacharyya et al. \cite{c29,c30}, which are also within the 2+1-flavor setup. The Stefan–Boltzmann (SB) limit for $v_s^2$ and $\frac{p}{\epsilon}$ is flavor independent and corresponds to the value, $v_s^2=\frac{P}{\epsilon}=1/3$.}
	\label{Therm3}
\end{figure*}
Fig. \ref{Therm3} (a) shows the squared speed of sound $v_s^2$, which decreases sharply near $T_c$. This dip reflects the softening of the EoS: pressure grows more slowly with energy density as the medium undergoes deconfinement. Beyond the transition, $v_s^2$ increases gradually and approaches the conformal limit of 1/3 at high temperatures. Panel (b) presents the ratio $P/\epsilon$ which displays similar behaviour, remaining below the Stefan–Boltzmann limit but after transition it rises, steadily with $T/T_c$. These illustrations also reveal a contrasting behaviour compared to the previously discussed quantities. In both cases, the QPM-II curve lies above the QPM-I curve up to a certain point ($T/T_c\approx1.0$ for $v_s^2$ and at $T/T_c\approx2.0$ for $P/\epsilon$), after which both converge. This behaviour arises because at higher temperatures $m_{eff}$ makes the quasiparticles heavier and less sensitive to temperature. As a result, it suppresses the energy density $\epsilon$ (and consequently the specific heat $C_V$) slightly more than it suppresses the pressure P (and the entropy density $s$). Thus, our findings in Fig. \ref{Therm3} (a) show strong agreement with both lattice QCD data \cite{c25} and the work of Bhattacharyya et al. \cite{c30} throughout the transition region. By contrast, in Fig. \ref{Therm3} (b), our results agree closely with lattice QCD data \cite{c27,c28} and the work of Bhattacharyya et al. \cite{c30} only at higher temperatures, specifically for $T/T_c\geq1.3$.
\\
Fig. \ref{Therm4} (a) compares $v_s^2$ and $P/\epsilon$. Both quantities exhibit a gradual increase with temperature and tend toward their respective conformal limits at higher $T/T_c$, indicating the expected approach of QCD matter toward conformal behaviour in the deconfined regime.
In Fig. \ref{Therm4} (b), the comparison of scaled specific heat ($C_V$) and energy density ($\epsilon$) with temperature can be seen. The specific heat is observed to be above energy density through the temperature range. Whereas, at higher temperatures both quantities converge and approach their respective conformal limits.
\\
Overall, both QPM-I and QPM-II models successfully reproduce the thermodynamics behaviour across the transition, displaying only mild differences in their prediction near $T_c$. These features include the rapid increase in pressure and entropy density, the peak in specific heat, the dip in sound speed, and the eventual convergence of all quantities toward the conformal limit at high temperatures. This reflects the thermodynamically self consistent dynamics of the quasiparticle description. In the high temperature region, our results appear slightly lower than the compared lattice QCD data. This is primarily due to the absence of strange quarks in our framework. As we utilize a 2-flavor model ($N_f$=2), the system inherently possesses fewer active degrees of freedom and thus a lower Stefan-Boltzmann limit than the 2+1-flavor lattice results (except for observables such as the speed of sound squared and $P/\epsilon$ whose limits are flavor independent). The suppression of specific heat also leads to a mild overestimation of the speed of sound squared. Regarding the chiral sector, its absence mainly affects quantities sensitive to the transition region such as the trace anomaly. However, at high temperatures where chiral symmetry is effectively restored, its influence becomes negligible. Therefore, the dominant source of deviation at high temperature is the reduced flavor content of the model. Furthermore, differences between QPM-I and QPM-II become more apparent at intermediate to high temperatures where a transient inversion in $v_s^2$ and $P/\epsilon$ reflects their distinct mass implementations. The strong medium induced effects on $m_q$ surpass those on $m_{eff}$ and makes it more sensitive at higher temperatures. Whereas, the residual interactions in  $m_{eff}$ tends to make it less sensitive to temperature after transition. At sufficiently high temperatures, both mass schemes behave very much similar and produce almost indistinguishable results. Therefore, the results converge smoothly to the standard PNJL predictions and are consistent with the expected high-temperature limit of QCD thermodynamics \cite{c1,c25,c27}.
Finally, although certain thermodynamic quantities such as specific heat in Fig. \ref{Therm2} (b) show apparent agreement with lattice data at low temperatures. This agreement should be interpreted while considering that the relevant degrees of freedom are hadronic and are not fully captured in the present framework. It is therefore primarily qualitative and serves an illustrative role rather than a quantitative validation.
\\
Above all, our two variants quasiparticle treatment (QPM-I and QPM-II) provide an important information of QGP's thermodynamics near and above the critical temperature for 2-flavor scheme under the suitable conditions at zero chemical potential ($\mu_0$=0) and zero magnetic field ($B$=0). This proposed effective PNJL+QPM extension situates within the broader landscape of current research and underscores the motivation for combining the two frameworks to study the QGP thermodynamics. Using the current baseline theoretical framework, future investigations incorporating heavy flavor, chemical potential, and/or magnetic fields are expected to open many challenging areas in the field of high-energy heavy-ion collisions not only in the high temperature sector but also important in the high density region.

\section{Conclusion}
\label{Section:5}
In this work, we have successfully investigated the thermodynamic properties of the quark-gluon plasma by applying a quasiparticle approach to the 2-flavor effective PNJL model. 
The calculated thermodynamic observables are found to be in close agreement with lattice QCD data across the temperature range of the crossover transition. 
\\
A key finding of our work is that both QPM-I and QPM-II formulations of the quasiparticle model, when integrated into the PNJL framework, produce results that align well with each other and also with the LQCD data. This confirms that the quasiparticle approach offers a robust and reliable alternative to the traditional PNJL models for describing the QGP thermodynamics. Both models, QPM-I and QPM-II complement each other as well and do not invalidate the calculation of EoS. 
\\
The implementation of a quasiparticle description within the PNJL model represents a crucial theoretical advance in the study of hot QCD matter. By incorporating the medium effects through the temperature dependent masses, the effective PNJL framework successfully resolves the major deficiencies of conventional mean-field approaches, particularly in the high-temperature EoS.

Finally, the results achieved from Model-I and Model-II variants across all thermodynamic observables validates the robustness of core quasiparticle mechanism in order to explore QCD phase diagram. The QPM extension provides a phenomenological approach of the effective PNJL model with temperature dependent quark masses, in which medium induced effects are incorporated within the PNJL framework. This formulation enables a consistent description of the QGP thermodynamics, particularly in the deconfined region near and above the transition temperature where quasiparticle behaviour dominates. 
\\
Within this regime, the model effectively captures the leading interaction effects and reproduces the overall thermodynamic trends. At the same time, the present formulation does not include an explicit dynamical treatment of the chiral condensate and is restricted to a 2-flavor system which limits its quantitative agreement with full 2+1-flavor lattice QCD, especially for observables sensitive to flavor content and transition dynamics. These features define the scope of applicability of the model rather than a deficiency and extensions incorporating additional flavors along with a dynamical chiral sector are expected to further improve the quantitative accuracy and broaden its scope. Future investigations may extend this framework to include higher-order perturbations, improved flavor dependence, finite chemical potential, and external magnetic fields to extend the applicability of the model and enable a more comprehensive exploration of the QCD phase diagram.


%
%

\end{document}